\documentclass{aa}
\usepackage{epsfig}
\begin{document}

\thesaurus{06(08.19.04; 13.07.01)}
\title{Upper limits to low energy $\bar\nu_\mathrm{e}$ flux from GRB990705}


\author{
M.Aglietta\inst{14}, E.D.Alyea\inst{7}, P.Antonioli\inst{1},
G.Badino\inst{14}, G.Bari\inst{1}, M.Basile\inst{1},
V.S.Berezinsky\inst{9}, F.Bersani\inst{1}, M.Bertaina\inst{14}, 
R.Bertoni\inst{14},
G.Bruni\inst{1}, G.Cara Romeo\inst{1},
C.Castagnoli\inst{14}, A.Castellina\inst{14}, A.Chiavassa\inst{14},
J.A.Chinellato\inst{3}, 
L.Cifarelli\inst{1}, F.Cindolo\inst{1},
A.Contin\inst{1}, V.L.Dadykin\inst{9},
L.G.Dos Santos\inst{3},
R.I.Enikeev\inst{9}, W.Fulgione\inst{14},
P.Galeotti\inst{14},
P.L.Ghia\inst{14}, P.Giusti\inst{1}, 
F.Grianti\inst{1},
G.Iacobucci\inst{1},
E.Kemp\inst{3}, F.F.Khalchukov\inst{9}, E.V.Korolkova\inst{9},
P.V.Korchaguin\inst{9}, V.B.Korchaguin\inst{9}, V.A.Kudryavtsev\inst{9},
M.Luvisetto\inst{1}, 
A.S.Malguin\inst{9},
T.Massam\inst{1},
N.Mengotti Silva\inst{3},
C.Morello\inst{14},
R.Nania\inst{1},
G.Navarra\inst{14},
L.Periale\inst{14}, A.Pesci\inst{1}, P.Picchi\inst{14},
I.A.Pless\inst{8}, V.G.Ryasny\inst{9},
O.G.Ryazhskaya\inst{9}, O.Saavedra\inst{14}, K.Saitoh\inst{13},
G.Sartorelli\inst{1}, 
M.Selvi\inst{1},
N.Taborgna\inst{5},
N.Takahashi\inst{12},
V.P.Talochkin\inst{9},
G.C.Trinchero\inst{14}, S.Tsuji\inst{10}, A.Turtelli\inst{3},
P.Vallania\inst{14}, 
S.Vernetto\inst{14},
C.Vigorito\inst{14},
L.Votano\inst{4}, T.Wada\inst{10},
R.Weinstein\inst{6}, M.Widgoff\inst{2},
V.F.Yakushev\inst{9}, I.Yamamoto\inst{11},
G.T.Zatsepin\inst{9}, A.Zichichi\inst{1}
}

\offprints{W. Fulgione Istituto di Cosmo-Geofisica corso Fiume 4, 
10133-Torino (I)}
\mail{FULGIONE@TO.INFN.IT}

\institute{
University of Bologna and INFN-Bologna, Italy
\and Brown University, Providence, USA
\and University of Campinas, Campinas, Brazil
\and INFN-LNF, Frascati, Italy
\and INFN-LNGS, Assergi, Italy
\and University of Houston, Houston, USA
\and Indiana University, Bloomington, USA
\and Massachusetts Institute of Technology, Cambridge, USA
\and Institute for Nuclear Research, Russian Academy of
Sciences, Moscow, Russia
\and Okayama University, Okayama, Japan
\and Okayama University of Science, Okayama, Japan
\and Hirosaki University, Hirosaki, Japan
\and Ashikaga Institute of Technology, Ashikaga, Japan
\and Institute of Cosmo-Geophysics, CNR, Torino, University
of Torino and INFN-Torino, Italy
}

\date{Submitted June 5, 2000; accepted xx, xxxx}

\titlerunning{$\bar\nu_e$ from GRB990705}

\authorrunning{LVD Collaboration}

\maketitle

\begin{abstract}

The detection of Gamma Ray Burst GRB990705
on 1999, July 5.66765 UT,
pointing to the Large Magellanic Clouds, suggested the search for
a possible neutrino counterpart, both in coincidence with and slightly
before (or after) the photon burst.

We exploited such a possibility by means of the LVD
neutrino telescope (National Gran Sasso Laboratory, Italy), which
has the capability to study low-energy cosmic neutrinos.
No evidence for any neutrino signal, over a wide range of time durations,
has been found, at the occurrence of GRB990705.
Due to the lack of information about both the source distance
and its emission spectrum, the results of the search are expressed in terms 
of upper limits, at the Earth, 
to the $\bar{\nu}_\mathrm{e}$ flux $\cdot$ cross-section,
integrated over different time durations,
$\int \int \Phi_{\bar \nu_\mathrm{e}}\sigma dE dt$.

Moreover, assuming thermal $\bar\nu_\mathrm{e}$ spectra at the source, 
upper limits to the $\bar\nu_\mathrm{e}$ flux, integrated over time duration, 
for different spectral temperatures, are 
obtained.
Based on these limits and on the expectations for $\nu$ emission
from collapsing astrophysical objects, the occurrence 
of a gravitational stellar
collapse can be excluded up to a distance $r \approx 50$ kpc, 
in the case of time coincidence with GRB990705, 
and $r \approx 20$ kpc, for the 24 hours preceding it.

\end{abstract}

\section{Introduction}

Gamma Ray Burst GRB990705 was
detected on 1999, July 5.66765 UT, by the BeppoSAX Gamma-Ray Burst Monitor, 
and localized by the BeppoSAX
Wide Field Camera (Celidonio et al., 1999). 
It was promptly noted (Djorgovski et al., 1999) 
that its position, in 
projection, corresponded to the outskirts of the Large Magellanic Cloud 
(LMC), and
it was suggested that,
if the burst was indeed located in the LMC or its halo, a search for a 
neutrino signal, coincident with, or just prior to the GRB, would be 
quite interesting.

At the time of GRB990705, the LVD neutrino observatory, 
located in the Gran Sasso
underground Laboratory, Italy, was regularly taking data,
with active scintillator mass $M=573$ tons.
The main purpose of the telescope is the search for neutrinos from 
gravitational stellar collapses in the Galaxy.

On July 19$^{\mathrm{th}}$ 1999, 
the result of a preliminary analysis of the LVD data 
recorded during 48 hours around the time of GRB990705 was reported (Fulgione,
1999), and
the absence of a neutrino signal, 
that would be expected from a gravitational stellar 
collapse in our Galaxy, was established (no additional results from
other neutrino observatories were reported).

The search for low-energy neutrinos possibly associated to GRBs is indeed
of interest, especially
in view of the recent observational evidence linking (some) GRBs 
and supernovae (see, e.g., Galama et al., 1998, Bloom et al., 1999, 
Reichart, 1999). 
Many recent widely
discussed models of the sources of GRBs involve 
the core collapse of massive stars 
(see, e.g., Woosley, 1993, Paczynski,
1998, Mac Fayden \& Woosley, 1999, Khokhlov et al., 1999,
Wheeler et al., 1999):
in this scenario the neutrino emission could be associated to the cooling 
phase of the collapsed object,
the time separation between the neutrino and gamma signals 
depending on the time
necessary to transfer energy from the central engine, which emits thermal
$\nu$, to the outer region, emitting high energy photons.

It is clear that the possibility of detecting neutrinos 
correlated to GRBs depends on the distance of the associated source: 
even if it appears established that most of them lie at cosmological distances
(Metzger et al., 1997), there is evidence, for 
at least one of GRBs, to be related to a supernovae event in the local universe
(Tinney et al., 1998). In particular, from the study of the
afterglow of GRB990705 (Masetti et al., 2000), 
although an extragalactic origin might be
supported, the association with LMC cannot be ruled out.

Consequently, 
a more careful analysis of the LVD data in correspondence of
GRB990705 has 
been performed, to search for 
weaker neutrino signals, not only in coincidence with, but also
preceding\footnote{
By analogy with SN explosions modelling, where few hours are required
for the shock to reach the star envelope and give rise to the sudden increase
of luminosity, a similar time gap can be assumed, between neutrinos and
high-energy $\gamma$-rays. 
}
and even shortly following it.

The paper is planned as follows: 
in Sect.2 we briefly describe the LVD detector, and we explain the structure
of the data.
In Sect.3 we present the results of the analysis: a search 
for a $\bar{\nu}_\mathrm{e}$ signal 
coincident in time with GRB990705 has been performed.
Moreover, a time interval spanning from 24 hrs preceding 
the burst up to 10 minutes later, has been scanned, searching for any 
non-statistical fluctuation of the background.
For sake of completeness, a wider interval, since 10 days before
to 1 day after the event, has been investigated.
We conclude in Sect.4, discussing the results
in terms of upper limits to the $\bar \nu_\mathrm{e}$ 
flux possibly associated to the GRB,
under the hypothesis of thermal neutrino energy 
spectrum at the source, 
and comparing such limits with the expectations from 
existing models on $\nu$ emission from
collapsing objects.  

\section{The LVD Experiment and the Data}

The Large Volume Detector (LVD) in the Gran Sasso Underground 
Laboratory, Italy, consists of an array of 840  
scintillator counters, 1.5  m$^3$ each, 
interleaved by
streamer tubes, arranged in a compact and modular 
geometry (see Aglietta et al., 1992, for a more
detailed description), with an active scintillator mass $M=1000$ tons. 
The experiment has been taking data, under different larger
configurations, since 1991 (at the time of GRB990705, the active mass was
$M=573$ tons).

The main purpose of the telescope is the detection of neutrinos from 
gravitational stellar collapses in the Galaxy,
mainly through the absorption interaction
$\bar \nu_\mathrm{e} \mathrm{p,e^+ n}$.
This reaction is observed in LVD counters through two
detectable signals:
the prompt signal due to the $\mathrm{e}^+$ 
(detectable energy 
$E_\mathrm{d} \simeq E_{\bar\nu_\mathrm{e}}-1.8$ MeV $+ 2\mathrm{m_e} 
\mathrm{c}^2$),
followed, with a mean delay  $\Delta t \simeq 200~\mu \mathrm{s}$, 
by the signal from
the $\mathrm{n p,d} \gamma$ capture ($E_{\gamma} = 2.2$ MeV).

Counters can be considered as divided into two subsets:
external, i.e. those directly exposed to the rock radioactivity,
which operate at energy threshold $E_{\mathrm{th}}\simeq 7$ MeV,
and inner (core), operating at $E_{\mathrm{th}}\simeq 4$ MeV.

In the search for antineutrino interactions 
($\bar\nu_\mathrm{e} \mathrm{p,e^+ n}$), 
raw data are processed in order to reject muons, and filtered on the basis
of the prompt pulse ($\mathrm{e}^+$) energy release 
and of the presence of delayed low energy signals (n capture).  

We define three classes of data:
\begin{itemize}
\item class A: pulses with $E_\mathrm{d} \geq 7$ MeV ($M=573$ tons);
\item class B: pulses with $E_\mathrm{d} \geq 7$ MeV,
followed by a delayed ($\Delta t \leq 600~ \mu \mathrm{s}$) 
low energy pulse in the 
same counter ($M=573$ tons);
\item class C: pulses detected by core scintillators 
($E_\mathrm{d} \geq 4$ MeV), 
followed by a delayed low energy pulse in the same counter
($M=256$ tons).
\end{itemize}

The average efficiency for n detection is 
$\bar \epsilon_\mathrm{n}\simeq 60\%$ 
for the core and $\bar \epsilon_\mathrm{n}\simeq 50\%$ for the whole detector.

\section{The Analysis} 

\begin{table*}[htb]
\caption[]{Number of events ($N_\mathrm{d}$) 
detected in coincidence with GRB990705,
for different duration of the time window ($\delta t$),
compared with the expectations from background.}
\begin{center}
\begin{tabular}{c c c c c c c }
\hline
Number of events & $ \delta t = 1$ s & $\delta t = 5$ s & $\delta t = 10$ s 
& $\delta t = 20$ s &$\delta t = 50$ s
&$\delta t = 100$ s \\
\hline
Observed: class A& 0 & 0  & 0 & 1 & 7 & 12 \\
$<N_{\mathrm{bk}}>$& 0.15 & 0.7 & 1.5 & 2.9 & 7.3 & 14.6 \\  
\hline
Observed: class B& 0 & 0  & 0 & 0 & 0 & 0 \\
$<N_{\mathrm{bk}}>$& 0.03 & 0.1 & 0.3 & 0.5 & 1.3 & 2.6 \\  
\hline
Observed: class C& 0 & 0  & 0 & 0 & 0 & 1 \\
$<N_{\mathrm{bk}}>$& 0.02 & 0.1 & 0.2 & 0.4 & 1.0 & 2.0 \\  
\hline
\end{tabular}
\label{mark}
\end{center}
\end{table*}

\begin{figure}
\mbox{\epsfig{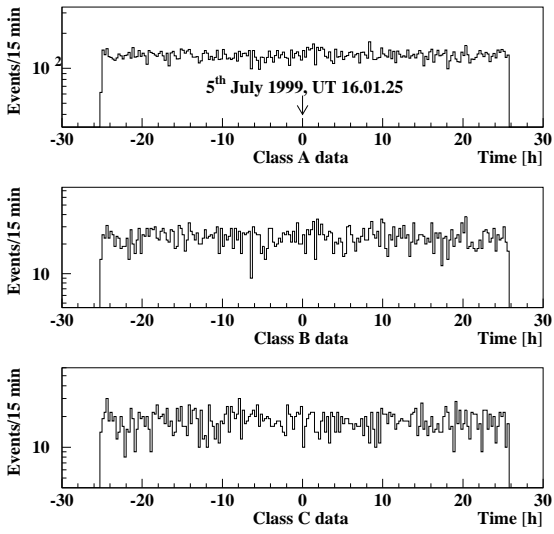}}   
\caption{Counting rates in the $48$ hours 
time window centered on the GRB990705.}
\label{fig:l1}
\vspace{-0.5cm}
\end{figure}

As a first step, the detector performance has been checked,
by studying the counting rate behavior during 48 hours around the time of 
GRB990705.
The number of events, detected every 15 minutes,
is shown in Fig.\ref{fig:l1}, for the three classes of data
defined in Sect.2: the stability of the counting rate, always 
within statistical
fluctuations, confirms the reliability
of the detector.

\subsection{In coincidence with GRB990705}

The search for a signal in time coincidence
with GRB990705 has been performed by comparing the number of
signals ($N_\mathrm{d}$), recorded during time windows having different 
duration $\delta t$, 
centered on the GRB time, 
with the average number of signals expected from background, 
$<N_{\mathrm{bk}}>$.
The value of $<N_{\mathrm{bk}}>$ 
has been evaluated by using the experimental rate in the
24 hours data after the GRB time (to avoid the contamination due to a possible
signal): the resulting statistical error is in any case $<3$\%.
 
Results are summarized in Table 1, for $\delta t = 1,5,10,20,50$ and 
$100$ s.
The differences between the number of detected pulses and expectations from 
background, for all event classes, are well within the statistical 
fluctuations.

No evidence for a $\bar\nu_\mathrm{e}$ 
signal coincident with GRB990705 appears 
from this analysis.

\subsection{Preceding (or following) GRB990705}

\begin{table*}[htb] 
\begin{center}
\begin{tabular}{ c c c c c } \hline
& \multicolumn{2}{c}{coincidence}  & \multicolumn{2}{c}{24 hour preceding}\\
\hline
$\delta t$ [s] & 
$\int\limits^{\delta t}_{0} dt
\int\limits^\infty_{5}
\frac{d^2\phi_\nu}{dEdt} \sigma(E) dE$ 
& 
$\int\limits^{\delta t}_{0} dt
\int\limits^\infty_{8}
\frac{d^2\phi_\nu}{dEdt} \sigma(E) dE$ 
&
$\int\limits^{\delta t}_{0} dt
\int\limits^\infty_{5}
\frac{d^2\phi_\nu}{dEdt} \sigma(E) dE$ 
& 
$\int\limits^{\delta t}_{0} dt
\int\limits^\infty_{8}
\frac{d^2\phi_\nu}{dEdt} \sigma(E) dE$\\
\hline
1 & $1.7 \cdot 10^{-31}$ & $4.3 \cdot 10^{-32}$ & $5.9 \cdot 10^{-31}$ &
$1.9 \cdot 10^{-31}$\\
5 & $1.7 \cdot 10^{-31}$ & $4.3 \cdot 10^{-32}$ & $5.9 \cdot 10^{-31}$ &
$2.4 \cdot 10^{-31}$\\
10 & $1.7 \cdot 10^{-31}$ & $4.3 \cdot 10^{-32}$ & $7.4 \cdot 10^{-31}$ &
$2.8 \cdot 10^{-31}$\\ 
20 & $1.7 \cdot 10^{-31}$ & $7.5 \cdot 10^{-32}$ & $8.1 \cdot 10^{-31}$ &
$3.5 \cdot 10^{-31}$\\
50 & $1.7 \cdot 10^{-31}$ & $8.6 \cdot 10^{-32}$ & $9.6 \cdot 10^{-31}$ &
$5.2 \cdot 10^{-31}$\\
100 & $2.9 \cdot 10^{-31}$ & $8.6 \cdot 10^{-32}$ & $1.1 \cdot 10^{-30}$ &
$6.0 \cdot 10^{-31}$\\
\hline  
\end{tabular}
\vspace{0.3cm}\\
{Tab.2: Upper limits (90\% c.l.) to the $\bar\nu_\mathrm{e}$ 
flux $\cdot$ cross-section, at the Earth, integrated over different 
time intervals.}
\end{center}
\end{table*}

The search for a possible $\nu$ burst 
has been extended to from 24 hours before GRB990705 occurrence
to 10 minutes after, for a total time $T=1450$ min.

\begin{figure}
\mbox{\epsfig{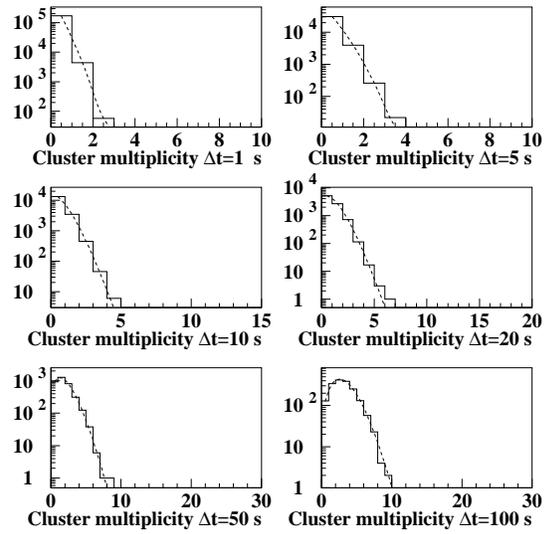}}   
\caption{Distributions of cluster multiplicity for events of class B 
detected during the 24 hours preceding GRB990705.
Dashed curves represent expectations from Poissonian background.}
\label{fig:l21}
\vspace{-0.5cm}
\end{figure}

The interval of interest has been divided into 
$N_{\delta t} = 2 \cdot \frac{T}{\delta t }$ intervals of duration 
$\delta t$,
each one starting at the middle of the previous one.
The multiplicity distributions of clusters (number of events within each 
interval of duration $\delta t$) have been studied for the three 
classes of data, defined in Sect.2, 
and for $\delta t = 1,5,10,20,50$,$100$ s, and they have been
compared with the expectations from Poissonian fluctuations
of the background.
In Fig.\ref{fig:l21}, we report, as an example, 
the result of the data analysis for class B events.

The agreement between data and expectations confirms the detector stability, 
allowing to state that there is no evidence for any detectable $\nu$ signal 
during the considered period.

For sake of completeness, the same analysis has been applied to the data 
collected since 240 hours preceding the GRB, up to 24 hrs later. 
Also in this case, the data are
in total agreement with the expectations from statistical fluctuations of the
background. 

\section{Results and Discussion}               


The number of expected $\bar\nu_\mathrm{e}$ interactions, $N_{\mathrm{ev}}$, 
in a time interval $\delta t$, due to a 
pulsed $\bar\nu_\mathrm{e}$ emission, is defined as:
$$N_{\mathrm{ev}} = M  \cdot N_\mathrm{p} 
\cdot \epsilon \int\limits^{\delta t}_{0} dt
\int\limits^\infty_{E_{\mathrm{min}}}
\frac{d^2\phi_{\bar\nu_\mathrm{e}}}{dE_{\bar\nu_\mathrm{e}} dt} 
\sigma(E_{\bar\nu_\mathrm{e}}) dE_{\bar\nu_\mathrm{e}}$$
where $\epsilon$ is the detection efficiency, $M$ [ton] is
the active scintillator mass, 
$N_\mathrm{p}$ is the number of free protons per scintillator ton, 
$\sigma(E_{\bar\nu_\mathrm{e}})$ is the neutrino interaction 
cross section (Vogel, 1984) and 
$\frac{d^2 \phi_{\bar\nu_\mathrm{e}}}{dE_{\bar\nu_\mathrm{e}} dt}$ is the
differential neutrino flux at the Earth. 

In the absence of any information on the source distance and
its emission spectrum, we can express the results of the search in 
terms of upper limits to the flux $\cdot$ cross-section, integrated over
the time duration, at the Earth:
$ \int dt \int \frac{d^2\phi}{dE dt} \sigma dE$.

These limits, calculated at 90\% c.l., are reported in Table 2,
for various 
burst duration $\delta t$, and they
are expressed in number of interactions per target proton.

%
%
%
%

\begin{figure}
\mbox{\epsfig{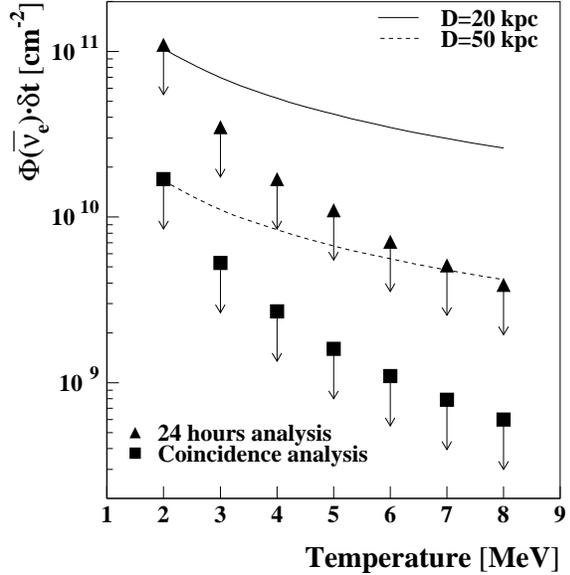}}   
\caption{Upper limits ($90 \%$ c.l.) to the time integrated 
$\bar\nu_\mathrm{e}$ flux,
at the Earth, for thermal 
$\bar \nu_\mathrm{e}$ spectra and $\delta t \leq 10$ s, 
compared with expectations
for different source distances.}
\label{fig:l2}
\vspace{-0.5cm}
\end{figure}

Any hypothesis on the $\bar\nu_\mathrm{e}$ source spectrum 
leads to a limit to the time
integrated $\bar\nu_\mathrm{e}$ flux at the Earth.
Assuming a thermal spectrum, constant during 
the emission interval $\delta t$, i.e.:
$$\frac {d\Phi_{{\bar\nu_\mathrm{e}}}}{dE_{\bar\nu_\mathrm{e}}} \propto \frac 
{(\frac{E_{\bar\nu_\mathrm{e}}}{T_{\bar \nu_\mathrm{e}}})^2}
{1+exp(- \frac{E_{\bar\nu_\mathrm{e}}}{T_{\bar \nu_\mathrm{e}}})}$$
upper limits to the time integrated $\bar\nu_\mathrm{e}$ flux
are obtained, as a function of the neutrinosphere 
emission temperature $T_{\bar \nu_\mathrm{e}}$ [MeV].
These results are shown in Fig.\ref{fig:l2}, for burst duration 
$\delta t \leq 10$ s.

Most theoretical models on the 
$\bar\nu_\mathrm{e}$ emission from gravitational stellar 
collapses (Burrows, 1992) predict that the neutron star binding 
energy,
$E_\mathrm{b} \approx 3 \cdot 10^{53}$ erg, 
is emitted in neutrinos of every flavour 
(energy equipartition) with thermal energy spectra, during a time interval 
$\delta t \approx 10$ s.
The corresponding $\bar\nu_\mathrm{e}$ 
fluxes at the Earth, calculated, under the
approximation of isotropical emission and
pure Fermi-Dirac spectrum, for two different source 
distances: $50$ kpc (i.e., corresponding to the LMC\footnote
{One can compare these results with the neutrino flux observed from 
SN1987A, which was definitely located in the LMC. 
According to the combined analysis of the events detected by the 
KamiokandeII and IMB detectors (Jegerlehner et al., 1996), which 
yields a total emitted energy $E_\mathrm{b} = 3.4 \cdot 10^{53}$ erg
and a $\bar\nu_\mathrm{e}$ 
spectral temperature $T_{\bar \nu_\mathrm{e}} = 3.6$ MeV,
the resulting $\bar\nu_\mathrm{e}$ flux at the Earth, integrated over time,
is $\Phi(\bar\nu_\mathrm{e})\cdot \delta t \sim 9. \cdot 10^{9} 
\mathrm {cm}^{-2}$
}
) and $20$ kpc 
(i.e., corresponding to the outskirts of our Galaxy), 
are reported in Fig.\ref{fig:l2} and
are compared with the results of the burst search.  

The occurrence of a gravitational stellar collapse, 
with $\bar \nu_\mathrm{e}$ emitted
in the temperature range $T_{\bar\nu_\mathrm{e}} > 2$ MeV, 
can then be excluded within a region of radius $r\approx50$ kpc, 
in the case of time 
coincidence with the GRB990705 event, and $r\approx20$ kpc, 
for the 24 hours preceding the GRB time\footnote{
A possible effect of neutrino mixing on the signal from a gravitational
stellar
collapse would result in the merging of the energy spectra of neutrinos of 
different flavours.
Because we are dealing with electron antineutrinos, which are characterized
by a spectral temperature lower then the one of $\bar\nu_{\mu}$ and 
$\bar\nu_{\tau}$, neutrino oscillation effects would lead 
to a hardening of the 
$\bar\nu_\mathrm{e}$ spectrum 
and, after all, to an increase of the $\bar\nu_\mathrm{e}$
detection probability.
Therefore, excluding oscillations into sterile neutrinos, the 
limits obtained in this work would 
remain valid even in the case of neutrino mixing.}.

\begin{acknowledgements}                   
The authors wish to thank the director and 
the staff of the National
Gran Sasso Laboratories for their constant and valuable support.
W.F. and P.L.G. gratefully acknowledge a useful discussion with 
Francesco Vissani.
\end{acknowledgements}

\end{document}